# Applying machine learning techniques to improve user acceptance on ubiquitous environment

Djallel Bouneffouf

Djallel.bouneffouf@it-sudparis.eu

**Abstract.** Ubiquitous information access becomes more and more important nowadays and research is aimed at making it adapted to users. Our work consists in applying machine learning techniques in order to adapt the information access provided by ubiquitous systems to users when the system only knows the user social group, without knowing anything about the user's interest. The adaptation procedures associate actions to perceived situations of the user. Associations are based on feedback given by the user as a reaction to the behavior of the system. Our method brings a solution to some of the problems concerning the acceptance of the system by users when applying machine learning techniques to systems at the beginning of the interaction between the system and the user.

**Keywords:** Context awareness, Machine learning, User acceptance.

## 1  Introduction

There was a tremendous technological revolution during last years with the arising of mobile terminals (e.g. computers, mobile computers, mobile phones, Pocket PC, PDA) and mobile networks (GSM, 3G+, wireless networks, Bluetooth, etc.). Using these devices, information processing has been thoroughly integrated into everyday activities, yielding to a new paradigm called ubiquitous computing. In ubiquitous computing, the quality of an information system is conditioned by the access to the right information at the right time and place. Given this, it is important to consider the needs of users and their contextual situation when they access information in order to provide them relevant results.

The need for adaptation of an information system to the user context has been accentuated by the extensive development of mobile applications that provide a considerable amount of data of all types (images, texts, sounds, videos, etc.). It becomes thus crucial to help users by guiding them in their access to information.

Ubiquitous computing takes advantage from this observation: its aim is to create smart systems where devices are dynamically providing new human-machine interaction possibilities. Systems should be able to recommend information helping the user to fulfill his/her goal. The information given by the system depends on the user's situation. Possible situations and the associated actions reflect the user's work habits.

To provide the relevant information to users, the system has to acquire the necessary knowledge to be able to decide which information to give at which moment. Our research work aims at automatically adapting ubiquitous systems to users. At the first time of interaction between the system and the user, the system does not know anything about the user, his/her preferences and habits. Consequently, to start reasoning, the system may refer to the habits of the social group to which the user belongs, and then by the time the system personalizes its services to the user. The information about the social group of a user is also called demographic information.

In the remaining of this paper, we present in section 2 our research problem and objectives. Section 3 is dedicated to the state of the art. Then, in section 4, we describe the current ideas of our ongoing work, followed by some preliminary results (section 5). Finally, we conclude, giving directions for future work.

## 2  Research problem

Our work relates to research in the following two areas:

- User's interest evolution depending on the context: A major difficulty when applying machine learning techniques to adapt a system to the user is the evolution of his/her interest. The interest of the user may change with the time and the system has to be continuously adapted to this dynamic change using the user's context information to provide the relevant recommendation.

- User acceptance: The system needs to respect some criteria to be accepted by the user, like intelligibility, nonintrusive and, avoiding the cold-start problem. In the scope of our work, we try to avoid the cold-start problem: the system should start with background knowledge because, at the initial state, if the system behavior is incoherent, the user quickly refuses it.

In this research context, our goal is to create a ubiquitous recommender system based on information about the user's context. Knowing this context, the system offers the relevant information to the user. For example, the system can remind a meeting when the user is not in his/her office. Most of current work on ubiquitous recommendation pre-defines services and fires them in the suitable situation [6, 3]. Our system starts with an initial default behavior defined by the social group to which the user belongs, and adapts it progressively to its particular users. The default behavior makes the system ready-to-use and the learning is a life-long process. In the beginning, the system is only acceptable but, with the time, it gives more and more satisfying results.

## 3   State of the art

There are many of research works, which propose to recommend relevant information to users, using different machine learning techniques. For example, in [10] the authors use linear classifiers trained only on rating information for recommendation. In [2], the authors propose a hybrid recommender framework to recommend movies to users. In the content-based filtering part of this hybrid recommender, they get extra information about movies from web site view each movie as a text document. A Naive Bayes classifier is used for building user and item profiles, which can handle vectors of bags-of-words.

These approaches have good results when they start with an amount of user's experience. However, none of them tried to work having no background knowledge about the user in the beginning.

Reinforcement Learning (RL) is an algorithm, which does not need a previous experience to start working. In [9], RL is applied to a recommendation system for the automatic acquisition of the context model in a ubiquitous environment. However, their recommendation system starts with a set of actions concerning the user which are predefined by experts.

In our work, we want to create a generic system, avoiding the intervention of experts because, in one hand, experts are not sure of the interest of the user, maybe defining wrong ideas about him; in the other hand, not always an expert is available.

To create a generic system avoiding the problem of cold-start, we propose to use demographic information to provide the relevant information to the user. In [7] the authors also use demographic information about users and items for providing more accurate prediction for users, but their system does not follow the user interest evolution.

## 4   Proposition

Our analysis of the state of the art allows us to say that the process of selecting relevant information to users, following their interest evolution, can be modeled as a process of RL. Moreover, Collaborative Filtering (CF) has been very successful [7] to avoid the problem of the cold-start using demographic information.

Our main idea is to start with an initial default behavior, defined by the social group to which the user belongs applying the CF. This behavior is normally acceptable to all users of the social network. Afterward, this behavior will be personalized trough the RL learning process.

### 4.1 Collaborative filtering

CF is built on the assumption that the best way to find interesting content is to find other people who have similar interests and then recommend items that those similar users like [1]. A CF recommender system works as follows. Given a set of transactions *D*, where each transaction *T* is of the form <*id*, *item*, *rating*>, a recommender model *M* is produced. Each item is represented by a categorical value while the rating is a numerical value in a given scale (e.g. each item is a movie rated with 1 to 5 stars). Such a model *M* can produce a list of *top-N* recommended items, and corresponding predicted ratings, from a given set of known ratings [4]. In many situations, ratings are not explicit. For example, if we want to recommend Web pages to a Web site visitor, we can use the set of pages he or she visited, assigning those pages an implicit rate of one, and zero to all the other pages.

In terms of CF, three major classes of algorithms exist (*Memory-based, Model-based Hybrid-based)* [1, 4]. At the moment, in our work, we use the simplest of them which is the memory-based CF.
*Memory-based*: the whole set of transactions is stored and is used by the recommender model. These algorithms employ a notion of distance to find a set of users, known as neighbors, who tend to agree with the target user. The preferences of neighbors are then combined to produce a prediction or *top-N* recommendation for the active user.

### 4.2 Reinforcement learning and the q-learning algorithm

RL is a computational approach of Machine Learning where an agent tries to maximize the total amount of reward it receives when interacting with a complex and uncertain environment [8]. A learning agent is modeled as a Markov decision process defined by *(S, A, R, P)*, where: *S* and *A* are finite sets of states and actions, resp.; $R : S \times A \rightarrow R$ is the immediate reward function and $P : S \times A \times S \rightarrow [0, 1]$ is the stochastic Markovian transition function. The agent constructs an optimal Markovian policy $\pi : S \rightarrow A$ that maximizes the expected sum of future discounted rewards over an infinite horizon. We define $Q^*(s, a)$, the value of taking action *a* in state *s* under a policy. The Q-learning algorithm allows computing an approximation of $Q^*$, independently from the policy being followed, if *R* and *P* are known.

### 4.3 The combination of collaborative filtering and reinforcement learning

The first step of the Q-Learning algorithm, called "action selection", is a compromise between the use of knowledge (we choose the action that maximizes the reward) and exploring the state space by choosing an action *a* a priori less interesting in terms of reward. In the Q-Learning algorithm, it is said that, for every state *s*, action *a = Q (s)* is chosen according to the current policy. The choice of the action by the policy must ensure a balance between exploration and exploitation.

The exploitation is to choose the best action for the current state, thus exploiting the system's knowledge at the moment of recommendation. The exploration is to choose an action other than the best in order to test it, observe its consequences, and increase the knowledge of the recommendation system.

It makes sense to encourage exploration in the beginning, to quickly cover a large part of the state space, then by the time decrease the exploration and increase the exploitation of knowledge to maximize the obtained rewards.

There are several strategies to make the balance between exploration and exploitation. Here, we focus on two of them:

- **greedy:** This strategy is not really an exploration because it chooses always the best action from the Q-table:

$$a = \arg \max_{a \in A} Q(s, a)$$

- **ε-greedy:** This strategy adds some greedy exploration policy, choosing a random action at each step if the policy returns the greedy action (probability ε) or a random action probability *(1 - ε)*.

To give the Q-Learning the ability to use advices from other users sharing the same ideas, we propose to extend the -ε-greedy strategy of the Q-Learning algorithm with the ability to explore the knowledge of other users. In the -ε-greedy strategy of the exploration/exploitation functions, we replace the random action by an action that is selected by calculating the similarity of user profiles applying the CF algorithm. We call the new algorithm "CF-QL", for "CF for Q-Learning".

## 5 Preliminary ideas

Fig 1 summarizes the global mechanism of the recommender system. To detect the user's context, the recommender system receives events from sensor module (the phone of user for example). These events constitute the input of the recommendation system and launch the learning algorithm of the recommendation system. This last one allows choosing an action to be executed in the environment.

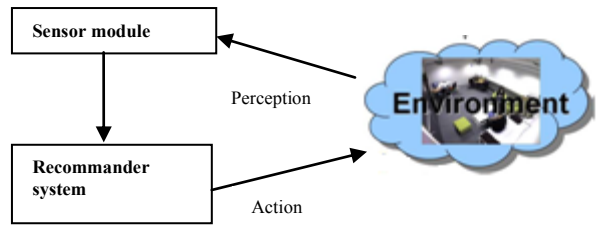

**Fig 1 :** Global mechanism of the interaction between system and environment

We consider the environment being composed of all context dimensions described in [5], namely cognitive, time, geographic, material, social, source document, qualitative, psychological, and demographic.

For the actual state of our work, the environment is replaced by a context simulator which executes scenarios by notifying the recommender system simulated events as real world sensors would do. And sensor module can only detect the time, the social, and the cognitive dimension of the context in the following way:
- The time is detected by the user mobile phone and the calendar of his/her company.
- The social group is predefined by his/her job.
- The cognitive dimension is detected by the interaction between the system and the user.

## 6 Preliminary results

We use an experimental platform to test our algorithms which is developed in C#.
For this test we used a very simple scenario: Given a company Nomalys, the set of marketing people can access to the most relevant data of their company via their mobile phone. Paul is a new sales representative of the company; he is integrating a team of ten marketing persons.
Regarding Paul's agenda, he has a meeting with a client in Paris at midday. When he arrives at his meeting, the system has to recommend him the client's register of complaints. To achieve this goal, the system only knows the habits of the ten colleagues of Paul.

In our experiments, the system is controlled by each of the previously presented algorithms: CF, Q-Learning, and CF-QL.

The experiment consists of testing 100 trials for each algorithm which start when the user connects to the system. During the trials, the system recommends resources to the user. We assume that if the user chooses one of the recommended resources (e.g. reads a document, opens a folder, etc.), it is considered as a good recommendation.

To evaluate each trial, we classify the possible results using the traditional Precision measure (percentage of good recommendations). Fig 2 shows the precision curves per 10 trial intervals for the three algorithms, where:

The experimental results show that CF-QL achieves better results than CF and Q-Learning. In general, the precision of CF-QL is equal or greater than the other ones except for trials number 20 and 50.

This small experiment gives an indication about the better performance of CF-QL algorithm with respect to the other ones.

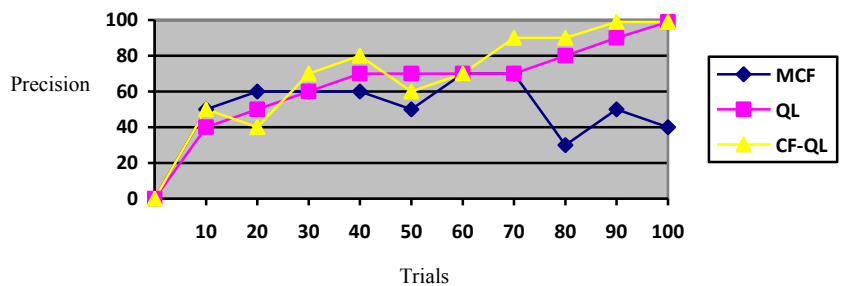

**Fig 2 :** Precision curves obtained for the CF, Q-Learning and the CF-QL algorithms

## 7 Conclusion

The aim of this PhD thesis is to investigate the adaptation and acceptance by the user of a recommendation system when the system does not know anything about the user and his habits in the beginning. The recommender system defines the observable situations and what actions should be executed in each situation in order to provide useful information to the user.

To achieve this goal, we start by applying the RL algorithm to solve the problem of the system starting without an amount of experience (cold-start), and the CF algorithm to initialize the learning process.
As future work, we intend to improve the adaptation of the recommendation system by integrating another context dimension as space and cognition dimension and, carrying out tests with real users.